\documentclass[prlaps,prl,twocolumn,twoside,showpacs,superscriptaddress]{revtex4}
\usepackage{graphicx}
\usepackage{epsfig}
\usepackage{color}
\begin{document}
\title{Slow dynamics in a turbulent von K\'arm\'an swirling flow }
\author{A. de la Torre}
\email{admonguio@alumni.unav.es}
\author{J. Burguete}
\email{javier@fisica.unav.es}
\affiliation{Departamento de F\'{\i}sica y Matem\'atica Aplicada\\
Universidad de Navarra, P.O.Box 177, E-31080 Pamplona, Spain}
\pacs{05.45.-a, 47.27.-i, 47.32.-y, 47.65.-d}
\textheight 250mm

\begin{abstract}
We present an experimental study of a turbulent von K\'arm\'an flow produced in a cylindrical container
using two propellers. The mean flow is stationary up to $Re = 10^4$, where a bifurcation takes place.
The new regime breaks some symmetries of the problem, and is time-dependent.
The axisymmetry is broken by the presence of equatorial vortices with a precession movement, being
the velocity of the vortices proportional to the Reynolds number.
The reflection symmetry through the equatorial plane is broken, and the shear layer of the mean flow appears displaced
from the equator. These two facts appear simultaneously. In the exact counterrotating case, a bistable regime appears between both mirrored solutions
and spontaneous reversals of the azimuthal velocity are registered. This evolution can be explained using a three-well potential model 
with additive noise. A regime of forced periodic response is observed when a very weak input signal is applied.
\end{abstract}

\maketitle

\textit{Introduction} -
Turbulent flows are ubiquitous in nature: ranging from small scales (heart valves, turbulent mixing) to 
very large scales (clouds, tornadoes, oceans, earth mantle, the sun and other 
astrophysical problems), turbulence is present in many applied and fundamental physical problems\cite{Frisch:1995}.
But, in spite of the attention it has received, there are still many open questions:
the emergence of coherent structures, as vortices, in fully developed turbulence 
or the rise of different bifurcations on the mean flow\cite{Holmes:1996}. 

Here we will analyze a particular configuration, the von K\'arm\'an swirling flow, where two different propellers are rotating 
inside a cylindrical cavity.  These flows have been studied analytically\cite{vonKarman:1921,Zandbergen:1987}, 
numerically \cite{Nore:2003,Nore:2006,Zhong:2006} and experimentally \cite{Ravelet:2005,Marie:2003,Nore:2005}.
Recently it has been shown that they can present multistability and 
memory effects\cite{Ravelet:2004}. Thus, this configuration is a natural candidate to study the instabilities that appear in
fully developed turbulence and the role played by symmetry breaking and coherent structures.
They have been also used in magnetohydrodynamics (MHD) experiments looking for the dynamo 
action with recent successful results \cite{Monchaux:2007} and with a very rich dynamics of the magnetic field \cite{Berhanu:2007}. 
Because of turbulence, the whole MHD problem can not be studied numerically: a usual approach is to deal with stationary mean 
flows in the so called kinematic dynamo scheme. 
Although these numerical studies have been used to predict thresholds of the dynamo action\cite{Gailitis:2000, Stieglitz:2001, Petrelis:2003}, 
real flows can present slow dynamics compared to the magnetic diffusion time that can not be neglected and should be taken into account in the numerical codes. 

In this paper we will focus on the slow dynamics that appear for very large $Re$ numbers.
The flow presents a non-trivial alternation between two states that break symmetries of the problem.
 This dynamics, which can be assimilated to a Langevin system 
with a classical exponential escape time and forced periodic response. 

\begin{figure}[!htb]
\epsfig{file=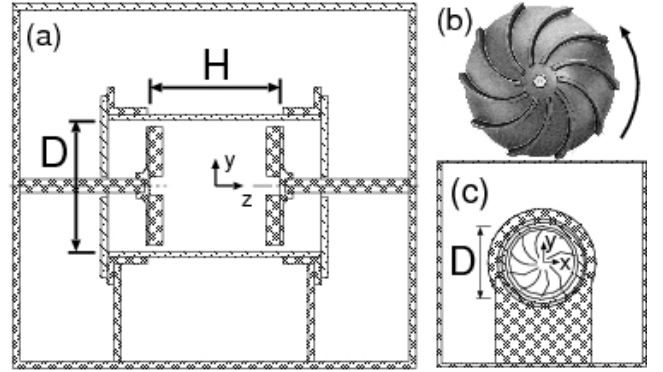, width=8.5cm}
\caption{Experimental setup. (a) Horizontal cylinder with the propellers, inside the tank. The north propeller is at $z=H/2$ and the south is at $z=-H/2$. 
(b) Photograph of the propeller. (c) Scheme of the south propeller viewed from the equatorial plane. 
The rotation sense with the convex side sets the azimuthal velocity as positive.}
\label{expsetup}
\end{figure}

\textit{Experimental setup} - 
The experimental volume (Fig.\ref{expsetup}) is a closed, horizontal cylinder whose diameter $D=2R=20\,$cm
 is fixed while the height $H$ can be modified continuously. 
Two propellers are placed at both ends, with radius $R_{prop}=8.75\,$cm and $10$ curved blades,
 each blade with a height of $2$ cm and a curvature radius of $5.0\,$cm.
Using the standard cylindrical coordinate system, the north $N$ (resp. south $S$) propeller placed at $z=H/2$ (resp. $z=-H/2$) 
has negative (resp. positive) azimuthal velocity.
\begin{figure*} 
\epsfig{file=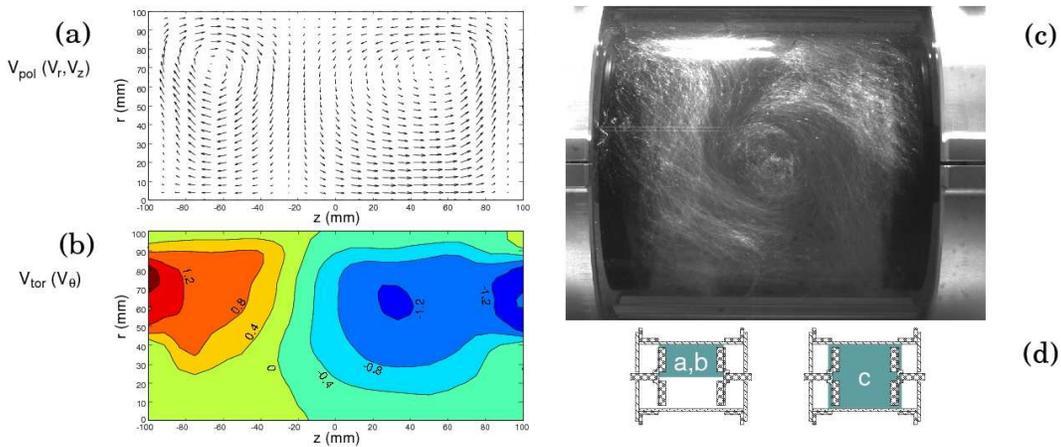, width=14cm} 
\caption{(Color online) (a) Stream-vectors $V_r$,$V_z$ in the plane $\theta=0$. (b) Contourplot of  $V_\theta$. 
The north propeller is on the right, the south propeller on the left. (c) Vortex visualized with air bubbles. 
(d) Gray zones indicating the regions where the figures (a,b,c) were obtained in the cell.}
\label{vort_mean}
\end{figure*}
The propellers are impelled by two independent motors of $3$ kW total power, allowing a rotation frequency in the range $f=0-20$ Hz. 
The frequency of the motors is controlled with a waveform function generator and a PID servo loop control.
The cylinder is placed inside a tank of $150$ l of volume in order to avoid optical problems and to assure the temperature stability. 
The fluid used is water at 21$^{\circ}$C.\par

The measurement of the velocity field is performed using a LDV system (with a chosen spatial resolution of $1\,$cm and 
temporal resolution up to $100$ kHz) and a PIV system (spatial resolution of $1$ mm and temporal resolution of $15$ Hz). 
The LDV system allows the measurement of two components of the velocity field (axial $v_z$ and azimuthal $v_{\theta}$), while 
the radial component ($v_r$) is obtained by mass conservation. 
The velocity field obtained in this way is consistent with the PIV measurements in an axial, horizontal plane ($\theta=0$).
Both techniques are based on the displacement of small particles ($14\,\mu$m, $\rho = 1.65\,$g/cm$^3$) inside the bulk of the fluid.\par

In our experimental setup three parameters can be modified. The first one is the Reynolds number $Re=R V_{prop} / \nu$, defined 
using the propeller's rim velocity $V_{prop}= 2 \pi R_{prop} f$. This number can be varied continuously in the range $Re\sim 10^3-10^6$. 
The second one is the aspect ratio of the experimental volume, $\Gamma = H/D $,  set to $1$ in this experiment. 
The last one is the frequency disymmetry, defined as $\Delta= (f_N - f_S)/(f_N + f_S)$, where $f_N$ and $f_S$ are 
the frequencies of each propeller, north and south. This parameter is varied in the range $\mid \Delta \mid < 0.1$.\par

\textit{Results} - 
For the $Re$ range explored, the flow is in fully-developed turbulence regime.
The mean flow $V=\langle v \rangle= \left( V_r,V_{\theta},V_z\right)$ represented in
Fig.\ref{vort_mean}(a,b) is obtained with the LDV system, averaging velocity series longer than $300$ times the period of the propeller. 
The measurement is done in the plane $\theta=0$ with $Re=3 \,\,10^5$ and $\Delta=0$. 
The flow is divided into two toroidal cells, each of them following its propeller (positive azimuthal velocity in the south, negative in the north). 
In each cell, the flow is aspired trough the axis towards the propellers, where it is ejected to the walls. 
The flow then returns along the cylinder's wall, approaching the axis near the equatorial plane. 
Using PIV measurements, the instantaneous field can be obtained, and no traces of the mean flow are observed. 
This is due to the high turbulence rate ($rms$ value over the mean value) which vary between $60-150\%$, 
depending on the spatial position and the velocity component measured.\par

This mean flow does not preserve the symmetry around the equator (a $\pi$ rotation around any axis in the $z=0$ plane, i.e. $R_{\pi}$ symmetry). 
In the case presented in figure \ref{vort_mean}(a,b) the cell near the north propeller is bigger than the south one.
The broken symmetry is recovered when the mirror state (south cell bigger than the north cell) is considered.
Each state (labeled as `north' $N$ or `south' $S$ depending on which cell is the \textit{dominant} one, i.e. bigger) 
are equally accessible when the system starts from rest.
This disymmetry is in contradiction with other works\cite{Ravelet:2005} in which the $R_{\pi}$ symmetry is observed (i.e., the 
frontier between the two rolls is always in the plane $z=0$), probably due to the presence of baffles or inner rings,
 which would enforce this condition. Other experiments at much slower Reynolds 
numbers~\cite{Nore:2005} ($Re\leq 600$) have shown that the axisymmetry can be broken producing near heteroclinic orbits,  but preserving 
the $R_{\pi}$ symmetry.\par

When larger tracers (air bubbles) are used, coherent structures as vortices (Fig.\ref{vort_mean}c) are visualized. These structures 
mostly relay in the dominant cell, so the azimuthal velocity of the vortices are directly related with the azimuthal velocity of the dominant cell.
These vortices have a characteristic size $D_{vortex} = 5\,$cm and they appear simultaneously to 
the disymmetry of the mean flow. 
Previous works in similar configurations \cite{Nore:2003} showed the formation of a static vortex in the equatorial 
plane ($z=0$) for much lower $Re$, inaccessible with the present configuration.\par
\begin{figure}[!htb]
\begin{center}
\epsfig{file=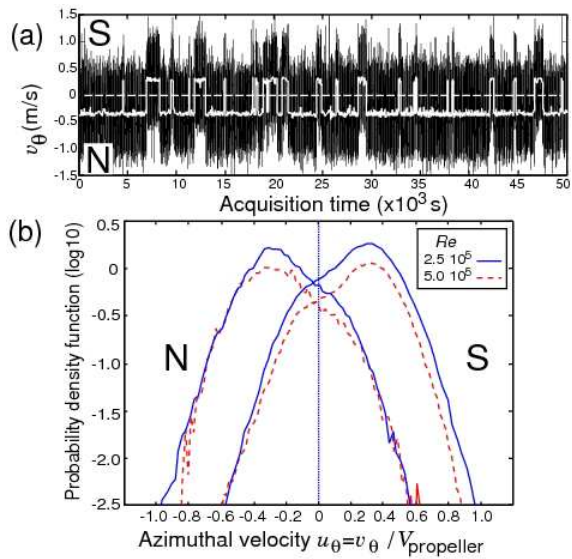, width=7.5cm}
\end{center}
\caption{\label{SerieInv} (a) Inversions of the azimuthal velocity at a point near the wall in $z=0$ for
 $\Delta=0$, $Re=2.5\,\,10^5$. Solid white line: filtered data using a low-pass filter with a cutoff frequency $f_{cut}=0.025$~Hz.(b) PDF of $u_\theta$ for each state 
($N$ or $S$) obtained for $\Delta=0$ and various $Re$.}%
\end{figure}

The state $N$ or $S$ of the system can be characterized using different variables. One is the position of the frontier $z_0$, defined as 
the $z$ position in where $V_{\theta}=0$ (in Fig.\ref{vort_mean}b, $z_0\sim -20$ mm). 
Another possibility is measuring $V_{\theta}^{eq}$, the mean azimuthal velocity at an equatorial point 
near the wall ($r=0.9R, z=0$) with the LDV system (in Fig.\ref{vort_mean}b, $V_{\theta}^{eq}\sim -0.4$ m/s).
 This mean velocity is stable in time and proportional to the propeller's rim velocity. For $Re<10^4$ we find that
the normalized azimuthal velocity ($U_{\theta}= V_{\theta}^{eq} / V_{prop}$) is nearly null so the mean flow is almost symmetric. 
As the $Re$ is increased the disymmetry becomes more notorious, until a plateau ($U_{\theta}= const$) is reached for $Re>10^5$.\par

For this range of $Re$ the system can spontaneously jump from one state to the other (\textit{inversions}). In Fig.\ref{SerieInv}(a) 
the instantaneous azimuthal velocity of the equator is plotted versus time ($t_{acq} =5\,\,10^4$ s $\sim 14$ h) for $Re=2.7 \,\,10^5$, $\Delta=0$. 
The typical transition time is about $10\,$s, while the time between inversions can vary from minutes to hours. 
These time scales are much slower than the period of the propeller ($t_{acq}=2.4\,\,10^5 \,\, T_{prop}$, with $T_{prop}=1 / f_{prop}=0.2$ s).

The probability density function (PDF) of these states can be computed breaking up the data series
in time intervals: a low pass filter is applied (fig.\ref{SerieInv}(a), white line)
that allows to differentiate between $N$ and $S$ and split the 
signal. 
The shape of these PDFs does not depend on the cutoff frequency
except for extreme values ($f_{cut}\rightarrow 0$ or $f_{cut}\rightarrow$ {sampling rate}).
Fig.\ref{SerieInv}(b) shows the PDF of the normalized instantaneous azimuthal 
velocity ($u_{\theta}= v_{\theta} / V_{prop} $) for the two states:
north (with negative most probable $u_{\theta}$) and south (with positive most probable velocity).
Each distribution ($p_N$ or $p_S$) is
described as the superposition of two gaussians:
\begin{eqnarray}
 p_{N,S}(u_{\theta})&=&G_0+ G_{N,S}=\frac{A_0}{\sqrt{2\pi}\sigma_0}\exp{\left( -\frac{u_{\theta}^2}{2\sigma_0^2}\right)}\nonumber\\
 &+&\frac{A_{N,S}}{\sqrt{2\pi}\sigma_{N,S}}\exp{\left(-\frac{(u_{\theta}-u_{N,S})^2}{2\sigma_{N,S}^2}\right)}
\label{gauss}
\end{eqnarray}
with $A_0+A_{N,S}=1$.
\begin{figure}[!htb]
\epsfig{file=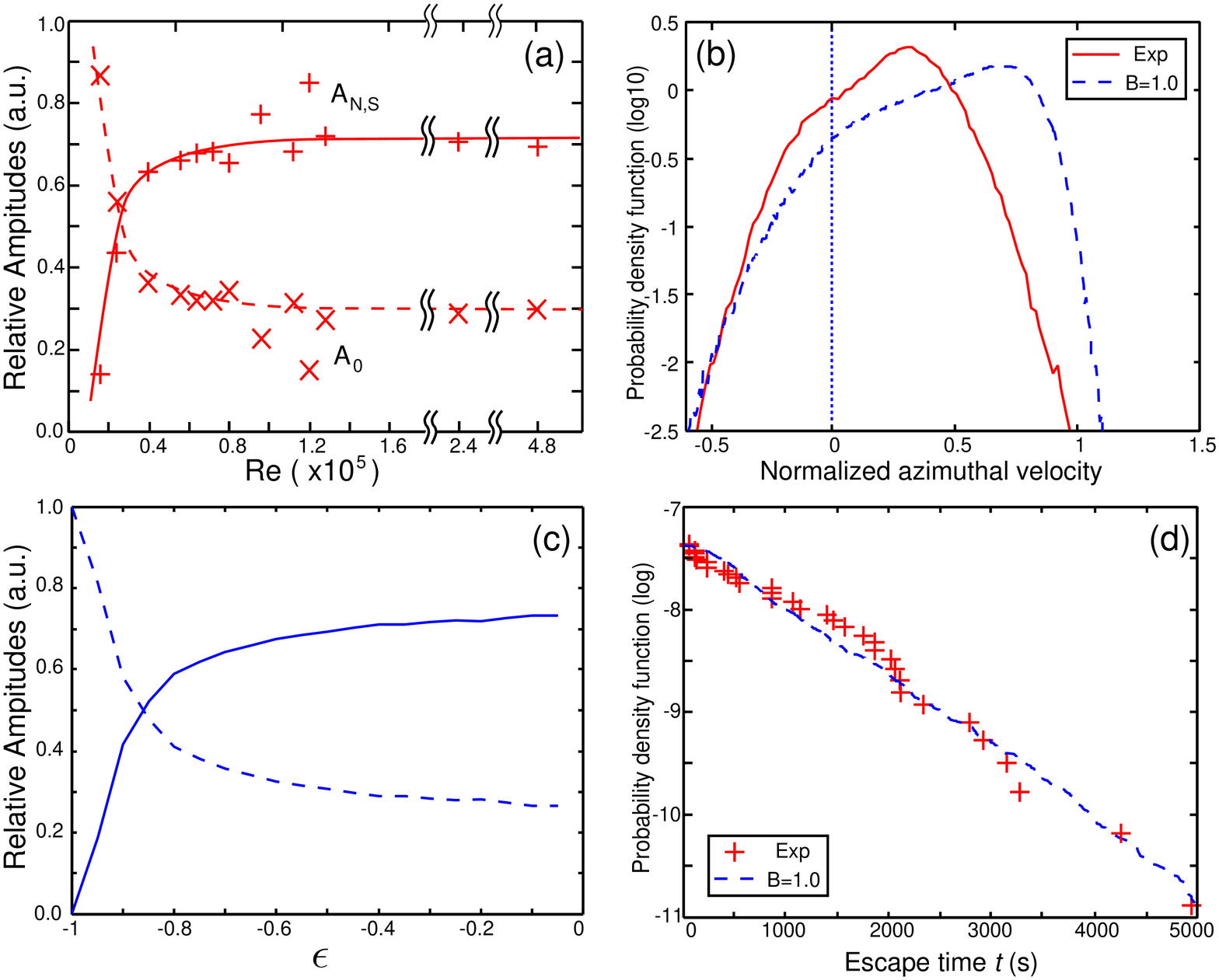,width=8.5cm}
\caption{(Color online) \label{amplitudes}%
(a) Experimental amplitude of the gaussians $A_0$ ({\color{red}$\times$,-\ -}) and $A_{N,S} $ ({\color{red}$+$,---}) 
	in Eq.\ref{gauss} $vs.$ $Re$. The lines are only plotted to indicate the trend and are not obtained from any fitting. 
(b) PDF of $u_{\theta}$ for the $S$ state for the experimental data series $Re=2.5\,10^5$ of Fig.\ref{SerieInv} (solid line) and numerical data (dashed line) 
	using the model~(eq.\ref{sysdin}) with $\epsilon=-0.05$, $g=2$, $B=1$.
(c) Numerical amplitudes of the $u_0$ ({\color{blue}---}) and $u_{N,S}$({\color{blue}-\ -}) $vs.$ $\epsilon$. 
	Each point corresponds to an ensemble average of 100 realizations. 
	The noise intensity increases linearly with $\epsilon$, in such a way that $B=0$ (resp. $1$)  when $\epsilon=-1$ (resp. $-0.05$). 
(d) PDF of the escape times for experimental data of Fig.~\ref{SerieInv} ({\color{red}$+$}) and for the numerical data presented 
	in (b) (dashed line, time unit $\tau_{sim}= 391\,$s).}
\end{figure}

Both distributions $p_N$ and $p_S$ share the same properties: one of the gaussians $G_0$ has zero mean, while the other $G_{N,S}$ is centered 
around a finite value ($|u_N|=|u_S|\neq 0$) that increases slightly with the $Re$ number. 
The amplitudes $A_0$ and $A_{N,S}$ have a stronger dependency on the $Re$ (Fig.\ref{amplitudes}a). 
For low $Re$ the PDF are nearly gaussians ($A_0 \simeq 1$)  while for large $Re$ (plateau) the non-symmetric 
gaussian becomes dominant $A_{N,S} \gg A_0 \neq 0 $.
The zero mean gaussian $G_0$ is due to residues of the symmetric flow and the other gaussian $G_{N,S}$ is related to the 
displacement of the vortices around the equator. 

According to this description, the system visit three different regions in phase space around $u_\theta =[0, u_N, u_S]$. 
A simple model based on a three well potential (one for 
the symmetric case $u_\theta=0$ and the two others for the asymmetric states $u_{N,S}$) will describe this dynamics:
\begin{equation}
\dot{u}_{\theta}=\epsilon u_{\theta}+ g\, u_{\theta}^3-u_{\theta}^5 +\sqrt{2B}\,\xi(t)
\label{sysdin}
\end{equation}
where $\xi(t)$ is a noise distribution with noise level $B$ (playing the role of the turbulence rate) and
 $g$ controls the relative depth of the potential wells.
Thus, the $N$ state (resp. $S$) will appear when the system is wandering between the wells $u_0$ and $u_N$ (resp. $u_0$ and $u_S$). 
The parameter $\epsilon$ 
is varied in the range ($ -g^2/4 <\epsilon <0$) where 
the three solutions $u_\theta= [0,\pm ( g/2 +(g^2/4 + \epsilon)^{0.5})^{0.5} ]$ are stable.

Different runs using an Euler-Maruyama scheme \cite{Kloeden:1999} were performed in order to recover 
the dynamics: for small $\epsilon$ the dynamics is confined to the region around $u_\theta=0$ whereas for $\epsilon\rightarrow 0$
the numerical evolution presents spontaneous inversions. In this later case the PDF of each state can
be computed and compared to experiments (fig.~\ref{amplitudes}.b).
The characteristic doubly bumped distribution is obtained, but in the numerical distribution the queues are not symmetric
due to the shape of the 6-$th$ order potential in the neighborhood of $u_S$.
The relative weight of each one of the solutions in the numerical PDFs can be calculated (fig.~\ref{amplitudes}.c) and compared 
with the experimental amplitudes of fig.~\ref{amplitudes}.a.

The distribution of the times the system stays in one
state (\textit{residence times}) follows an exponential decay law (Kramer's escape rate
\cite{Kramers:1940,Hanggi:1990,Chechkin:2005}):
\begin{equation}
\rho(t)= 1 / T_0 \exp\left(-t/T_0\right)
\end{equation}
where $T_0$ is related to the intensity of noise.
In figure \ref{amplitudes}.d we present the experimental residence time for the data of fig.~\ref{SerieInv}.a.
The experimental data have a characteristic time $T_0=1484\,$s $= 7020\, T_{prop} = 0.19\, \tau_\nu $, being $\tau_\nu=R^2/\nu$ the diffusion time scale.

As a consequence of the dynamics, a natural question that arises in this problem is the response of the experiment to an external forcing. 
When a sinusoidal modulation is applied to the 
frequency of one of the propellers, the rim velocity evolves as $V_{prop}(t)= V_{prop}^0 \left( 1 + 2\Delta_0 \cos (\omega t)\right) $
where $\Delta_0$ is the maximum frequency disymmetry.

\begin{figure}[!htb] 
\epsfig{file=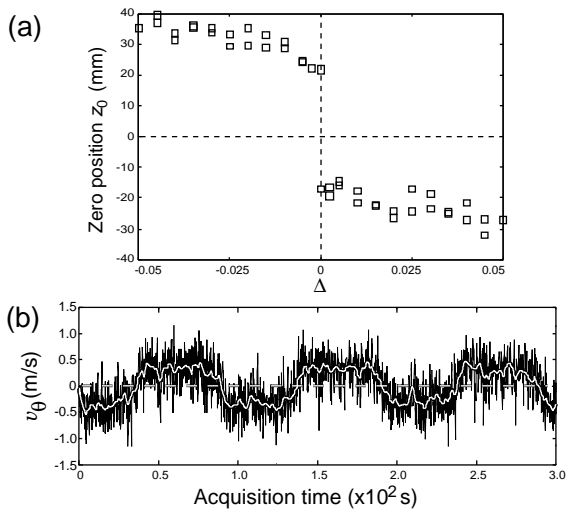, width=7.5cm}
\caption{\label{poscero} (a) Position of the frontier ($z_0$) varying the difference between the propellers' frequency ($\Delta$). 
The system presents a bistability at $\Delta =0$. (b) Temporal evolution of the instantaneous velocity of an equatorial point near the wall, with 
$\Delta=0.026\,\cos(0.02 \pi t)$, $Re=2.6\,10^5$. The system remains slaved to the input signal. The time scale is two orders of magnitude
smaller than that of figure~\ref{SerieInv}.a}
\end{figure}

A static forcing ($\omega=0$, $\Delta_0 \neq 0$) was applied to obtain the minimum amplitude
 needed to induce an inversion in the experiment.
In Fig.\ref{poscero}(a)  $z_0$ is represented $vs$ $\Delta$. 
A disimmetry of only $\Delta\sim 0.0025$ is needed to make the system jump between both states.
No inversions have been observed with $\Delta \ne 0$, so up to our precision and measurement times no hysteresis is found,
 and the system only shows a bistability in $\Delta=0$.
Further investigation will determine if $T_0 < \infty$ when $\Delta \ne 0$.\par

When the harmonic forcing is induced, preliminary results show how for low $f$ only a small amplitude in the input signal is 
needed to obtain the resonance of the system: Periodic inversions are observed with the same frequency of the forcing. 
Figure \ref{SerieInv}.b shows how the system is slaved to a signal with $\Delta_0=0.026$ and $f=10^{-2}\,$Hz.
In the range $10^{-2}> f / f_{prop} >10^{-3}$ it is necessary to increase $\Delta$ to observe the synchronism. 
For higher $f$, the system can not follow the forcing: the inertia of the flow limits the response 
time of the inversion $T_{res} \sim 20\,$s $=2.6 \,10^{-3}\, \tau_{\nu}$.
Further work experimental and numerical 
is under run in this direction.
\par


\textit{Conclusions} -
We have presented experimental evidence of a bifurcation in a turbulent system that breaks symmetries and produces slow dynamics. 
Two possible mirror states are equally accessible, with random (natural) or periodic (induced) inversions. 
The slow dynamics can be characterized using a very simple Langevin model in spite of being in a fully developed turbulence regime. 
Finally, when a very weak amplitude input modulation is applied to the system, a forced periodic response appears in the turbulent flow. 
One open question is either these natural inversions could be present in dynamo experiments for very long temporal series
($T_0 \simeq 1000$s) when $\Delta=0$.  \par

\textit{Acknowledgments} -
We are grateful to Jean Bragard, Iker Zuriguel and Diego Maza for fruitful discussions.
This work has been supported by the Spanish government (research projects FIS2004-06596-C02-01 and
UNAV05-33-001) and by the University of Navarra (PIUNA program).
One of us, AdlT, thanks the ``Asociaci\'on de Amigos'' for a post-graduate grant.


\begin{thebibliography}{16}

\bibitem{Frisch:1995}  U.~Frisch, \emph{Turbulence} (Cambridge Univ. Press, 1995).

\bibitem{Holmes:1996}  P.~Holmes, J.L.~Lumley, G.~Berkooz,
  \emph{Turbulence, Coherent Structures, Dynamical Systems and
  Symmetry} (Cambridge University Press, 1996).

\bibitem{vonKarman:1921} T.~von K\'arm\'an,  Z. Angew Math. Mech \textbf{1}, 233 (1921).

\bibitem{Zandbergen:1987} P.~Zandbergen, D.~Dijkstra, Ann. Rev. Fluid Mech. \textbf{19}, 465 (1987).

\bibitem{Zhong:2006} W.~Z. Shen,J.~Norensen, J.~Michelsen, Phys. Fluids \textbf{18},  064102 (2006).

\bibitem{Nore:2006} C.~Nore, L.~M. Witkowski, E.~Foucault, J.~Pecheux, O.~Daube, P.~Le~Quere, Phys. Fluids \textbf{18}, 054102 (2006).

\bibitem{Nore:2003} C.~Nore, L.~S. Tuckerman, O.~Daube, S.~Xin, J. Fluid Mech. \textbf{477}, 51 (2003).

\bibitem{Ravelet:2005} F.~Ravelet,  A.~Chiffaudel,  F.~Daviaud, J.~Leorat, Phys. Fluids \textbf{17},  117104 (2005).

\bibitem{Marie:2003} L.~Marie,  J.~Burguete,  F.~Daviaud,  J.~Leorat,  Eur. Phys. J B \textbf{33},  469 (2003).

\bibitem{Nore:2005} C.~Nore,  F.~Moisy, L.~Quartier,  Phys. Fluids \textbf{17},  064103 (2005).

\bibitem{Ravelet:2004} F.~Ravelet,  L.~Marie,  A.~Chiffaudel, F.~Daviaud,  Phys. Rev. Lett. \textbf{93}, 164501 (2004).

\bibitem{Monchaux:2007} R.~Monchaux  {\it et~al.}, Phys. Rev. Lett. \textbf{98}, 044502 (2007).

\bibitem{Berhanu:2007} M.~Berhanu {\it et al.}, Europhys Lett {\bf 77} (2007) 59001.

\bibitem{Gailitis:2000} A.~Gailitis {\it et al.}, Phys. Rev. Lett.  {\bf 84}, 4365 (2000).

\bibitem{Stieglitz:2001} R.~Stieglitz and U.~M\"uller, Phys. Fluids  {\bf 13} 561 (2001)

\bibitem{Petrelis:2003} F.~Petrelis {\it et al.} Phys. Rev. Lett. {\bf 90} 174501 (2003) 

\bibitem{Kloeden:1999} {{P.E.}~Kloeden}, {E.}~{Platen} {, \emph{Numerical Solution of Stochastic Differential Equations} (Springer, 1999).}

\bibitem{Kramers:1940} H.~Kramers, Physica A \textbf{7}, 284 (1940).

\bibitem{Hanggi:1990} {{P.}~{H\"anggi}},  {{P.}~{Talkner}},  {Rev. Mod. Phys.} \textbf{{62}},  {251} ({1990}).

\bibitem{Chechkin:2005} {{A.}~{Chechkin}},  {{V.}~{Gonchar}},  {{J.}~{Klafter}}, {{R.}~{Metzler}},  {Europhys. Lett.} \textbf{{72}},  {348} ({2005}).

\end{thebibliography}
\end{document}